\pdfoutput=1
\documentclass[runningheads]{llncs}

\usepackage[T1]{fontenc}
\usepackage{graphicx}
\usepackage{amsmath,amssymb,amsfonts}
\usepackage{caption}
\usepackage{listings}
\usepackage{booktabs}
\usepackage{textcomp}
\usepackage{tabularx}
\usepackage{float}
\usepackage{marvosym}

\begin{document}

\title{Influence of Prompt Engineering on Small Language Models for Guarded Query Routing}

\titlerunning{Small Language Models for Guarded Query Routing}

\author{Richard Šléher\inst{1}\orcidID{0009-0000-0069-969X} \and
William Brach\inst{1}\orcidID{0009-0002-0321-0321}(\Letter) \and
Kristián Košťál\inst{1}\orcidID{0000-0003-0679-4588} \and
Lukas Galke Poech\inst{2}\orcidID{0000-0001-6124-1092}}

\authorrunning{Šléher et al.}

\institute{Faculty of Informatics and Information Technologies, Slovak University of Technology in Bratislava, Bratislava, Slovakia\\
\email{\{xsleher,william.brach,kristian.kostal\}@stuba.sk} \and
Department of Mathematics and Computer Science (IMADA), University of Southern Denmark, Odense, Denmark\\
\email{galke@imada.sdu.dk}}

\maketitle

\begin{abstract}
We study the problem of guarded query routing, where we assume that a user query first meets a router that either determines the ideal endpoint for in-distribution queries, or rejects out-of-distribution queries that are potentially unsafe or out of the system's scope.
We investigate whether compact open-weight Small Language Models (SLMs) can jointly handle both tasks under latency constraints. We evaluate 22 models on GQR-Bench and score them with the harmonic mean of in-distribution and out-of-distribution accuracy. We find that mid-scale SLMs come close to frontier model routing quality at much lower latency. Still, many compact models fail because they do not reliably follow the required output format.
However, our results show that prompt optimization techniques enable SLMs to handle such cases gracefully -- without changing the models' weights. Moreover, few-shot prompt optimization raises Mistral~7B from 81.79 to 90.87 GQR-Score and lifts Qwen3.5~9B to 95.74, the best optimized score in our study and within 0.3 points of the strongest unoptimized larger model: Gemma~3~27B at 96.01. The bare DSPy signature, without in-context exemplars, is the most effective strategy for Granite~4~Tiny, raising its score from 54.29 to 83.05. These results show that prompt optimization is a useful first step for guarded query routing, while weaker models may still need weight-level adaptation or schema-aware training.

\end{abstract}

\section{Introduction}

In many LLM deployments, a user query does not go directly to a large model \cite{dekoninck2024unified}. It first passes through a router, which decides which expert should handle the request. In a guarded system, this first decision is also a safety decision, since the router should reject any query that is off-topic, adversarial, or outside the supported domains before downstream tools are invoked~\cite{b1,b35}. A guarded router, therefore, has to do two things at once. It must classify valid in-distribution (ID) queries, while rejecting out-of-distribution (OOD) inputs.

This makes the router important but also fragile. It sits on the critical path of every request, so its latency is paid every time the system is used. Its mistakes are also costly. An unsafe query may reach a downstream expert, while a legitimate query may be blocked. The problem is harder than ordinary intent classification because OOD queries can look similar to valid domain queries, and confidence scores often become unreliable under distribution shift~\cite{b22,b23}. LLM-based detectors inherit this problem and add their own sensitivity to prompt wording~\cite{b9,b10}. A practical router must therefore be fast, selective, and stable under near-OOD inputs~\cite{b19,b20}.

One simple design is to place a classifier in front of a separate guardrail. That design is easy to understand, but it incurs two costs per request and splits one semantic decision into two thresholds. The classifier decides where the query should go; the guardrail later decides whether it should have gone anywhere at all. If these two decisions disagree, errors compound. A single router with an explicit reject option avoids this split by treating dispatch and refusal as one calibrated decision~\cite{b24,b25}. The open question is whether a compact open-weight model can make this joint decision well enough while staying fast.

This paper makes two empirical contributions. First, we benchmark 22 open-weight models on GQR-Bench~\cite{b1} and show where they fall on the latency--efficacy trade-off. Mid-scale language models approach much larger models, while smaller models often reject OOD inputs but fail to commit to the correct ID label. Second, we study whether prompt-only optimization can repair this failure. On seven representative models, DSPy's Few-Shot optimization and GEPA show that simple in-context examples can recover ID accuracy by up to 57.6 percentage points without weight updates.

\section{Related Work}
\label{sec:related}

Guarded Query Routing (GQR) was recently formalized in~\cite{b1} as a joint routing and rejection problem. The router must send valid queries to one of several expert domains and reject queries that are off-topic, in another unsupported setting, or unsafe. GQR-Bench captures this setting with three ID domains (Law, Finance, Healthcare) and seven OOD splits. The original benchmark showed that efficient classifiers such as fastText~\cite{bojanowski2017enriching,joulin2017bag} and WideMLP~\cite{galke-scherp-2022-bag} offer a strong latency--efficacy trade-off, often running far faster than LLM routers. Our work starts with that result and asks a follow-up question: whether compact, open-weight LLMs can become better guarded routers merely by changing their prompts.

This question connects to several lines of work. Semantic routers use embeddings or keyword overlap to choose among routes~\cite{b21,b32}, but guarded routing also needs a reject option. The fields of out-of-distribution detection and abstention directly study the rejection problem~\cite{b22,b23,b24,b25,b38}, but often place less emphasis on ID classification. They show why confidence alone is risky under a distribution shift, especially when an OOD query shares words with a valid domain. These methods often assume a separate discriminative classifier. A compact generative model offers a different path, producing the route label and the reject label in one forward pass when it follows the output schema reliably.

SLMs are attractive here as they can be employed much more efficiently than frontier models while still performing well on classification tasks~\cite{b11,b12,galke2022we}. That matters because the router runs before every downstream LLM call~\cite{b19,b20,b21}. But the benefit comes with a catch: In GQR, the model must return one of a few exact labels, which entails a structured-output and instruction-following problem~\cite{b36,b37}.

This is why prompt engineering is a natural first tool. If the routing schema or rejection policy changes, a prompt can be inspected, changed, and rolled back without retraining the model. For a guarded router, that kind of lightweight adaptation is useful, since the system can be adjusted without changing weights or redeploying a new model.

Prompting methods range from manual templates to automated search~\cite{b26,b27,b28}. Few-shot prompting is especially relevant because examples can show the model what a valid routing answer looks like~\cite{b30,b31}. Automated systems such as MIPROv2 and GEPA go further by searching over instructions and demonstrations without changing model weights~\cite{b3,b29,b41}. These methods fit our setting well, as they do not come with additional resource requirements and allow rapid adaptation if, for example, a new domain is added.

Most prior routing work focuses on where to send a query, while most guardrail work focuses on whether a query is safe~\cite{b11,b19,b20,b21,b32,b34,b35,b39,b40}. GQR combines both questions into a single decision. The router must determine whether the query falls within the supported domains at all and, when it does, which expert should receive it. This leaves a gap. We still do not know how compact open-weight language models behave when routing and rejection are forced into one small label set. We study that gap directly. We keep the router structure fixed, score the joint decision with the GQR-Score~\cite{b1}, and study how far prompt adaptation can move compact models without weight updates.

\section{Setup}
\label{sec:setup}

\paragraph{Task:} Each query has one of four possible outcomes, namely Law, Finance, Healthcare, or a synthetic class for OOD examples. The first three are supported ID domains, while the last means that the query should be rejected. We run all evaluations on GQR-Bench~\cite{b1}, which combines ten public datasets into this single routing setup. Three splits provide the ID domains. Seven splits test rejection. Five contain unsafe or adversarial content (Jigsaw~\cite{borkan2019jigsaw}, OLID~\cite{zampieri2019olid}, HateXplain~\cite{mathew2021hatexplain}, dkhate~\cite{sigurbergsson2020dkhate}, and the Slovak-language TUKE SK set~\cite{hladek2023tukesk}), while two contain benign but unsupported topics: general web questions from Web Q and machine-learning questions from ML Q.

\paragraph{Evaluation Metrics:} A guarded router should not win by refusing everything, and it should not win by sending every query to an expert. We therefore report ID Accuracy, OOD Accuracy, and the GQR-Score from GQR-Bench~\cite{b1}. ID Accuracy is the macro-average over the three supported domains. OOD Accuracy is the size-weighted average over the seven rejection splits. GQR-Score, the harmonic mean of ID and OOD accuracy, is our headline metric because it rewards routers that do both jobs well. We also report Unsafe Avg., the size-weighted mean over the five unsafe OOD subsets, to isolate safety-critical refusal.

\subsection{Models and Prompt Optimization}
\label{subsec:models}

We evaluate 22 open-weight models spanning ten model families or release lines, namely Gemma~3~\cite{b7} (270M--27B), Gemma~4~E2B (5B), Llama~3~\cite{b5} (3B--70B), Qwen3~\cite{b4} (4B--14B), Qwen3.5 (0.8B--9B), Phi-4~\cite{b6} (14B), Mistral~\cite{b13} (7B), Granite~3.3~\cite{b14} (2B--8B), Granite~4~Tiny (7B), and GPT-OSS~\cite{b33} (20B). This diversity of architectures, training recipes, and parameter counts ensures that our findings on guarded routing do not depend on any single release.

From this pool, we select seven models for prompt optimization, namely Qwen3.5~9B, Qwen3.5~4B, Qwen3.5~2B, Qwen3.5~0.8B, Gemma~4~E2B~5B, Granite~4~Tiny~7B, and Mistral~7B. This subset spans four families and covers the 0.8B--9B parameter range where over-rejection under standard prompting is most pronounced. Rather than hand-tuning prompts, we employ the DSPy~\cite{b2} framework, which provides a stable and reproducible prompt optimization implementation. The router signature is displayed in Listing~\ref{lst:dspy-router-signature}. Each selected model is run under four prompting strategies. The first is \emph{standard prompting} with a fixed baseline template (Listing~\ref{lst:baseline-router-prompt}), used to isolate the effect of DSPy from the raw text prompts employed with the initial pool. The second is the \emph{DSPy baseline}: the bare \texttt{dspy.Predict(Classify)} program with no in-context exemplars, which isolates the effect of DSPy's signature-driven prompt formatting from any exemplar selection. The third is \emph{Few-Shot} optimization via \texttt{BootstrapFewShotWithRandomSearch}. This optimizer runs the unoptimized program on the training inputs, retains those whose traces produced the correct route as candidate in-context demonstrations, samples several alternative demonstration sets, and returns the single best-scoring program on a held-out validation slice; it selects exemplars by search and does not average outputs across runs. We use \texttt{max\_labeled\_demos=6} and \texttt{num\_candidate\_programs=4} on 30 training samples from GQR-Bench, deliberately below the DSPy defaults of 16 to yield a time-efficient configuration suited to single-machine exploration. The fourth is \emph{GEPA} evolutionary refinement~\cite{b3}, where we employ \texttt{gpt-5.4} as the teacher model; GEPA is initialized from the Few-Shot prompt and runs under the automatic \texttt{light} preset with 100 training and 30 validation samples drawn from GQR-Bench.

\clearpage
\begin{lstlisting}[caption={Baseline prompt used for LLM-as-a-Router under standard prompting}, label={lst:baseline-router-prompt}]
system_prompt = """You are a highly accurate text classifier. Your task is to categorize passages into one of four predefined domains. The ONLY valid categories are: law, finance, healthcare, and ood. Any passage that does not clearly belong to law, finance, or healthcare MUST be categorized as ood. You must respond with ONLY the category name, and nothing else. No explanations, no extra words."""

user_prompt = """Classify the following passage into one of the categories: law, finance, healthcare, or ood.
Passage:
{query}
Category:"""
\end{lstlisting}

\begin{lstlisting}[caption={DSPy routing signature used during prompt optimization}, label={lst:dspy-router-signature}]
class Classify(dspy.Signature):
    """
    You are a highly accurate text classifier. Your task is to categorize queries
    into one of four predefined domains. The ONLY valid categories are: law, finance, healthcare, ood
    Any query that does not clearly belong to the domains above MUST be categorized as ood.
    You must respond with ONLY the category name, and nothing else. No explanations, no extra words.
    """
    query: str = dspy.InputField(desc="The query to classify.")
    route: Literal["law", "finance", "healthcare", "ood"] = dspy.OutputField(desc="The predicted category. If the query does not clearly belong to category - law, finance, healthcare. Predict 'ood'.")
\end{lstlisting}

\subsection{Hardware and Inference Protocol}
\label{subsec:hardware}

All experiments ran on a single machine with two RTX 4090 GPUs, an AMD Threadripper PRO 7965WX (24 cores), and 256\,GB RAM. We serve models with vLLM~\cite{b18} using greedy decoding (temperature~0) and the default quantization for each model tag. Reported latency is the mean single-request inference time over the full test split, measured warm-start with one request in flight at a time, which matches the interactive routing setting where sub-second responses are expected. All results come from a single run and should be read as point estimates for this configuration.

\section{Experiments}

We examine whether open-weight SLMs can function as guarded query routers. We report results for standard prompting, DSPy-optimized prompting, and DSPy with GEPA refinement, using the GQR-Bench evaluation protocol.

\begin{table}[H]
\centering
\caption{Performance of evaluated open-weight models on GQR-Bench. We compare ID Accuracy, OOD Accuracy, and their harmonic GQR-Score. Strong guarded routers must both dispatch valid ID queries and reject OOD inputs. Key takeaways are that Gemma~3~27B is the strongest model overall (96.01 GQR), Qwen3.5~9B under Few-Shot gives the best optimized score (95.74 GQR), and Granite~4~Tiny illustrates the high-OOD/low-ID failure mode under standard prompting before the DSPy baseline raises it to 83.05 GQR.}
\label{tab:model-ood-results-full-big}
\resizebox{\textwidth}{!}{%
\begin{tabular}{lccccccc|c|cc|c}
\toprule
\textbf{Model} & \textbf{Jigsaw} & \textbf{OLID} & \textbf{HateXplain} & \textbf{dkhate} & \textbf{TUKE SK} & \textbf{Web Q} & \textbf{ML Q} & \textbf{Unsafe Avg.} & \textbf{ID Acc.} & \textbf{OOD Acc.} & \textbf{GQR-Score}  \\
\midrule
\multicolumn{12}{c}{\emph{Non LLM baselines}}\\
fastText & 74.46 & 61.51 & 54.46 & 74.77 & 83.11 & 70.37 & 63.28 & 69.66 & 95.80 & 68.85 & 80.12 \\
WideMLP ($\tau$=0.99) & 93.83 & 93.49 & 91.00 & 86.93 & 80.60 & 99.16 & 93.75 & 89.17 & 84.49 & 91.25 & \textbf{87.74} \\
\midrule
\multicolumn{12}{c}{\emph{Standard prompting}}\\
Llama~3.3~70B & 97.60 & 85.81 & 97.84 & 96.66 & 96.98 & 91.98 & 100.00 & 94.98 & 96.04 & 95.27 & 95.65 \\
Qwen3~30B & 98.88 & 91.74 & 99.07 & 96.66 & 96.14 & 91.14 & 99.22 & 96.50 & 94.33 & 96.12 & 95.22 \\
Gemma~3~27B & 97.39 & 93.84 & 99.36 & 98.78 & 98.54 & 94.49 & 99.22 & 97.58 & 94.69 & 97.37 & \textbf{96.01} \\
GPT-OSS~20B & 97.29 & 90.58 & 97.89 & 93.92 & 91.76 & 92.91 & 100.00 & 94.29 & 94.99 & 94.91 & 94.95 \\
Phi-4~14B & 95.08 & 80.47 & 96.71 & 92.71 & 88.74 & 89.42 & 93.75 & 90.74 & 96.03 & 90.98 & 93.44 \\
Qwen3~14B & 98.35 & 94.07 & 98.65 & 93.92 & 93.74 & 92.37 & 100.00 & 95.75 & 93.19 & 95.87 & 94.51 \\
Gemma~3~12B & 97.64 & 92.91 & 99.12 & 97.57 & 95.41 & 94.34 & 99.22 & 96.53 & 94.56 & 96.60 & 95.57 \\
Qwen3.5~9B & 99.10 & 90.93 & 99.02 & 97.87 & 96.25 & 93.80 & 100.00 & 96.63 & 92.48 & 96.71 & 94.55 \\
Granite~3.3~8B & 98.97 & 96.05 & 99.46 & 100.00 & 99.58 & 94.59 & 100.00 & 98.81 & 75.54 & 98.38 & 85.46 \\
Qwen3~8B & 99.60 & 98.60 & 99.71 & 99.09 & 99.58 & 92.96 & 99.22 & 99.32 & 84.46 & 98.40 & 90.89 \\
Mistral~7B & 99.35 & 97.56 & 99.98 & 99.70 & 99.17 & 99.26 & 100.00 & 99.15 & 69.53 & 99.29 & 81.79 \\
Granite~4~Tiny~7B & 99.91 & 100.00 & 100.00 & 100.00 & 99.90 & 98.47 & 100.00 & 99.96 & 37.29 & 99.75 & 54.29 \\
Gemma~4~E2B~5B & 97.92 & 89.30 & 98.23 & 94.53 & 95.52 & 92.72 & 96.88 & 95.10 & 92.63 & 95.01 & 93.81 \\
Qwen3.5~4B & 99.19 & 95.23 & 99.31 & 98.48 & 96.66 & 94.00 & 100.00 & 97.78 & 88.80 & 97.55 & 92.97 \\
Qwen3~4B & 99.28 & 97.33 & 99.44 & 97.57 & 97.60 & 86.71 & 98.44 & 98.24 & 76.99 & 96.62 & 85.70 \\
Gemma~3~4B & 98.38 & 94.42 & 99.49 & 97.87 & 98.64 & 97.88 & 100.00 & 97.76 & 85.04 & 98.10 & 91.11 \\
Llama~3.2~3B & 99.91 & 100.00 & 100.00 & 100.00 & 100.00 & 99.85 & 100.00 & 99.98 & 18.11 & 99.97 & 30.67 \\
Granite~3.3~2B & 99.94 & 99.07 & 100.00 & 99.70 & 99.69 & 96.95 & 100.00 & 99.68 & 54.42 & 99.33 & 70.32 \\
Qwen3.5~2B & 97.01 & 91.28 & 95.60 & 96.35 & 94.58 & 83.76 & 76.56 & 94.96 & 90.42 & 90.74 & 90.58 \\
Gemma~3~1B & 26.14 & 41.63 & 35.77 & 52.58 & 45.36 & 16.68 & 0.00 & 40.30 & 65.43 & 31.17 & 42.22 \\
Qwen3.5~0.8B & 63.66 & 38.95 & 58.92 & 57.45 & 54.85 & 21.51 & 44.53 & 54.77 & 79.73 & 48.55 & 60.35 \\
Gemma~3~270M & 10.36 & 5.12 & 3.22 & 1.82 & 0.83 & 1.72 & 8.59 & 4.27 & 50.28 & 4.52 & 8.30 \\
\midrule
\multicolumn{12}{c}{\emph{DSPy baseline}}\\
Qwen3.5~9B & 97.54 & 78.84 & 96.04 & 90.58 & 90.09 & 93.11 & 100.00 & 90.62 & 94.84 & 92.31 & \textbf{93.56} \\
Granite~4~Tiny~7B & 91.79 & 80.12 & 91.76 & 86.93 & 79.77 & 59.50 & 45.31 & 86.07 & 90.90 & 76.45 & 83.05 \\
Mistral~7B & 98.23 & 89.88 & 98.99 & 94.83 & 91.87 & 85.38 & 91.41 & 94.76 & 67.18 & 92.94 & 77.99 \\
Gemma~4~E2B~5B & 92.16 & 73.49 & 90.03 & 86.93 & 78.83 & 81.15 & 93.75 & 84.29 & 94.72 & 85.19 & 89.70 \\
Qwen3.5~4B & 98.01 & 88.26 & 97.62 & 96.96 & 94.99 & 94.64 & 100.00 & 95.17 & 89.44 & 95.78 & 92.51 \\
Qwen3.5~2B & 94.90 & 78.37 & 92.23 & 92.10 & 85.30 & 74.51 & 58.59 & 88.58 & 79.06 & 82.29 & 80.64 \\
Qwen3.5~0.8B & 99.97 & 99.42 & 99.97 & 100.00 & 99.90 & 97.64 & 100.00 & 99.85 & 11.22 & 99.56 & 20.17 \\
\midrule
\multicolumn{12}{c}{\emph{Few-Shot}}\\
Qwen3.5~9B & 99.13 & 93.02 & 98.97 & 97.26 & 96.25 & 95.13 & 99.22 & 96.93 & 94.52 & 97.00 & \textbf{95.74} \\
Granite~4~Tiny~7B & 83.39 & 70.93 & 83.03 & 65.65 & 53.91 & 55.56 & 57.81 & 71.38 & 94.88 & 67.18 & 78.66 \\
Mistral~7B & 98.16 & 91.98 & 98.10 & 93.92 & 91.24 & 95.67 & 89.06 & 94.68 & 87.93 & 94.02 & 90.87 \\
Gemma~4~E2B~5B & 97.45 & 86.28 & 95.97 & 95.14 & 94.68 & 89.62 & 94.53 & 93.90 & 95.02 & 93.38 & 94.19 \\
Qwen3.5~4B & 98.01 & 88.26 & 97.62 & 96.96 & 94.99 & 94.64 & 100.00 & 95.17 & 89.44 & 95.78 & 92.51 \\
Qwen3.5~2B & 65.81 & 59.42 & 74.76 & 80.85 & 71.43 & 40.94 & 1.56 & 70.45 & 93.78 & 56.40 & 70.43 \\
Qwen3.5~0.8B & 87.37 & 74.77 & 92.35 & 52.28 & 37.33 & 26.33 & 10.94 & 68.82 & 89.71 & 54.48 & 67.79 \\
\midrule
\multicolumn{12}{c}{\emph{GEPA}}\\
Qwen3.5~9B & 98.57 & 90.58 & 98.53 & 97.57 & 94.99 & 94.29 & 100.00 & 96.05 & 94.81 & 96.36 & \textbf{95.58} \\
Granite~4~Tiny~7B & 89.83 & 76.16 & 89.74 & 82.37 & 80.29 & 49.56 & 32.03 & 83.68 & 95.09 & 71.43 & 81.58 \\
Mistral~7B & 84.44 & 70.23 & 83.61 & 84.19 & 71.22 & 75.30 & 57.81 & 78.74 & 92.37 & 75.26 & 82.94 \\
Gemma~4~E2B~5B & 93.22 & 80.00 & 92.10 & 89.97 & 86.24 & 89.91 & 67.19 & 88.31 & 94.71 & 85.52 & 89.88 \\
Qwen3.5~4B & 98.48 & 92.44 & 98.26 & 96.66 & 94.79 & 97.54 & 100.00 & 96.13 & 87.08 & 96.88 & 91.72 \\
Qwen3.5~2B & 86.65 & 75.58 & 86.76 & 86.32 & 74.45 & 57.92 & 39.06 & 81.95 & 93.52 & 72.39 & 81.61 \\
Qwen3.5~0.8B & 99.78 & 98.95 & 99.95 & 100.00 & 99.58 & 94.59 & 100.00 & 99.65 & 32.49 & 98.98 & 48.92 \\
\bottomrule
\end{tabular}%
}
\end{table}

\subsection{Baseline Routing Performance}

\begin{figure}[htb]
\centering
\includegraphics[width=\columnwidth]{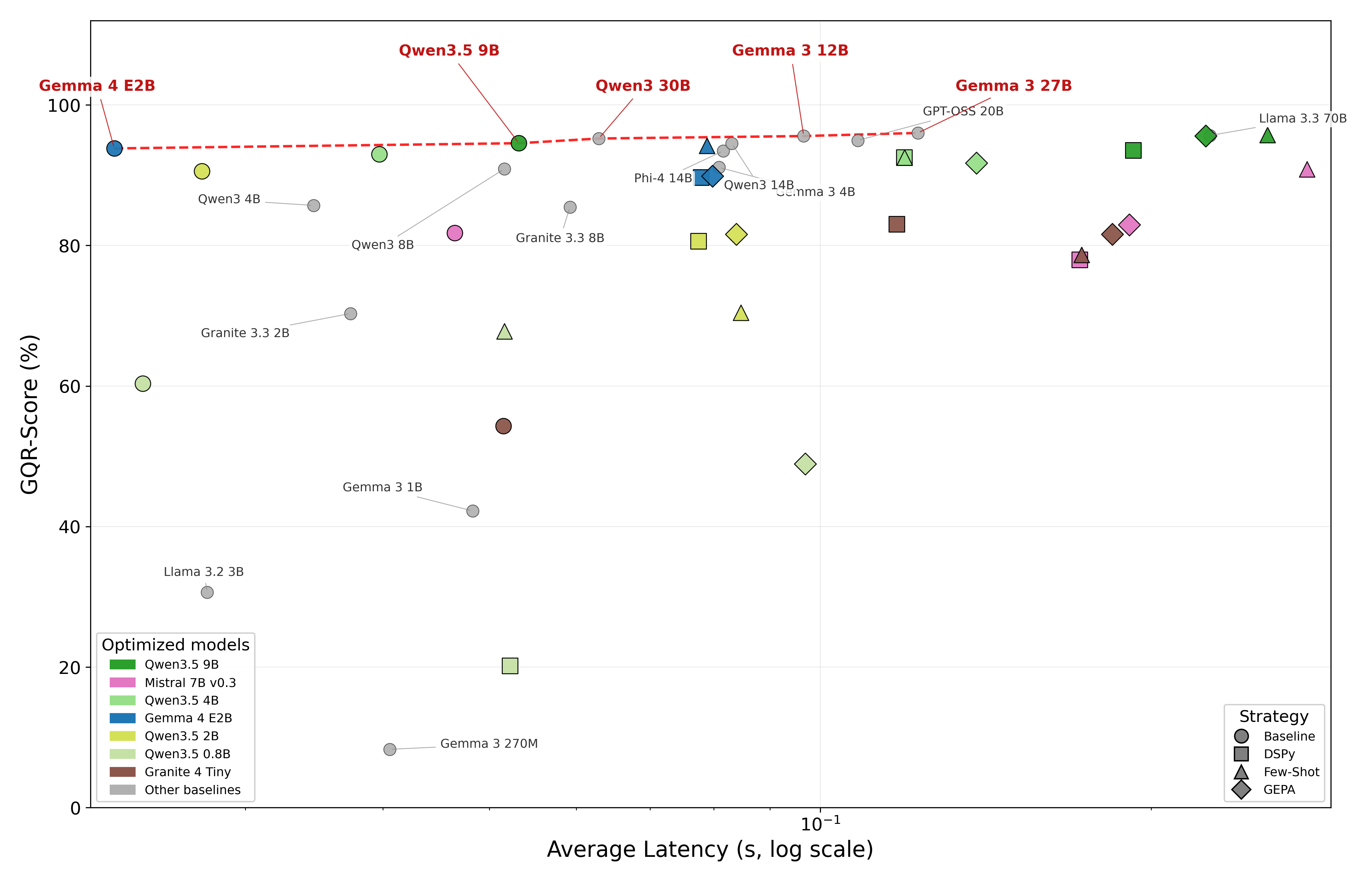}
\caption{Inference latency vs.\ GQR-Score trade-off of evaluated open-weight models on GQR-Bench. The horizontal axis shows inference latency on a logarithmic scale, and the vertical axis reports the GQR-Score. Red markers connected by a dashed line indicate the Pareto frontier, while gray markers are non-frontier models. The main interpretation is that several mid-scale SLMs occupy the useful upper-left region, where they approach large-model quality while remaining much cheaper to serve as per-query routers.}
\label{fig:tradeoff}
\end{figure}

The first result is that guarded routing does not simply reward the largest model. Under the standard prompt in Listing~\ref{lst:baseline-router-prompt}, GQR-Scores in Figure~\ref{fig:tradeoff} range from 8.30 (Gemma~3~270M) to 96.01 (Gemma~3~27B). The top three models, Gemma~3~27B (96.01), Llama~3.3~70B (95.65), and Gemma~3~12B (95.57), differ greatly in size but finish within 0.44 points of one another. This suggests that GQR-Bench saturates once a model is capable enough. The more important break appears below roughly 2B parameters, where models such as Gemma~3~270M (8.30) and Gemma~3~1B (42.22) no longer maintain a usable decision boundary.

A deeper analysis of the components of the GQR score reveals the main failure: Mid-scale models usually know when something should be rejected, but they often hesitate on valid ID queries. Granite~4~Tiny reaches 99.75\% OOD accuracy but only 37.29\% ID accuracy (GQR 54.29); Llama~3.2~3B reaches 99.97\% OOD accuracy, but only 18.11\% ID accuracy (GQR 30.67); and Mistral~7B reaches 99.29\% OOD accuracy but only 69.53\% ID accuracy (GQR 81.79). These models look safe because they reject almost everything, but a router that blocks valid requests is not useful. The pattern points to format-constrained instruction following~\cite{b36,b37}, not missing domain knowledge. The model often recognizes that a query belongs somewhere, but it does not reliably commit to one of the four allowed labels.

Across OOD splits, Web Q and OLID are typically the lowest-accuracy columns under standard prompting (Table~\ref{tab:model-ood-results-full-big}), indicating that benign but out-of-scope queries and offensive language remain the hardest rejection subtasks when the router must simultaneously maintain an ID routing signal.

\subsection{Effect of Prompt Optimization}

We then ask whether prompts can teach the router to commit to the right label. We apply the three DSPy~\cite{b2} strategies defined in Section~\ref{subsec:models} (DSPy baseline, Few-Shot via \texttt{BootstrapFewShotWithRandomSearch}, and GEPA~\cite{b3}) to seven models selected from the standard-prompting pool, namely Qwen3.5~9B, Qwen3.5~4B, Qwen3.5~2B, Qwen3.5~0.8B, Gemma~4~E2B~5B, Granite~4~Tiny~7B, and Mistral~7B. Table~\ref{tab:model-ood-results-full-big} reports every strategy model combination.

Few-Shot gives the best GQR-Score for four of the seven models (Qwen3.5~9B, Mistral~7B, Gemma~4~E2B, and Qwen3.5~0.8B). The DSPy baseline wins for Granite~4~Tiny, and standard prompting remains best for Qwen3.5~4B and Qwen3.5~2B. GEPA does not yield the best score for any model in this set. The largest gains occur when the model over-rejects under standard prompting. Granite~4~Tiny gains +28.8 GQR points under the DSPy baseline, while Qwen3.5~2B gains nothing because its standard-prompting score was already balanced.

\begin{figure}[t]
\centering
\includegraphics[width=\linewidth]{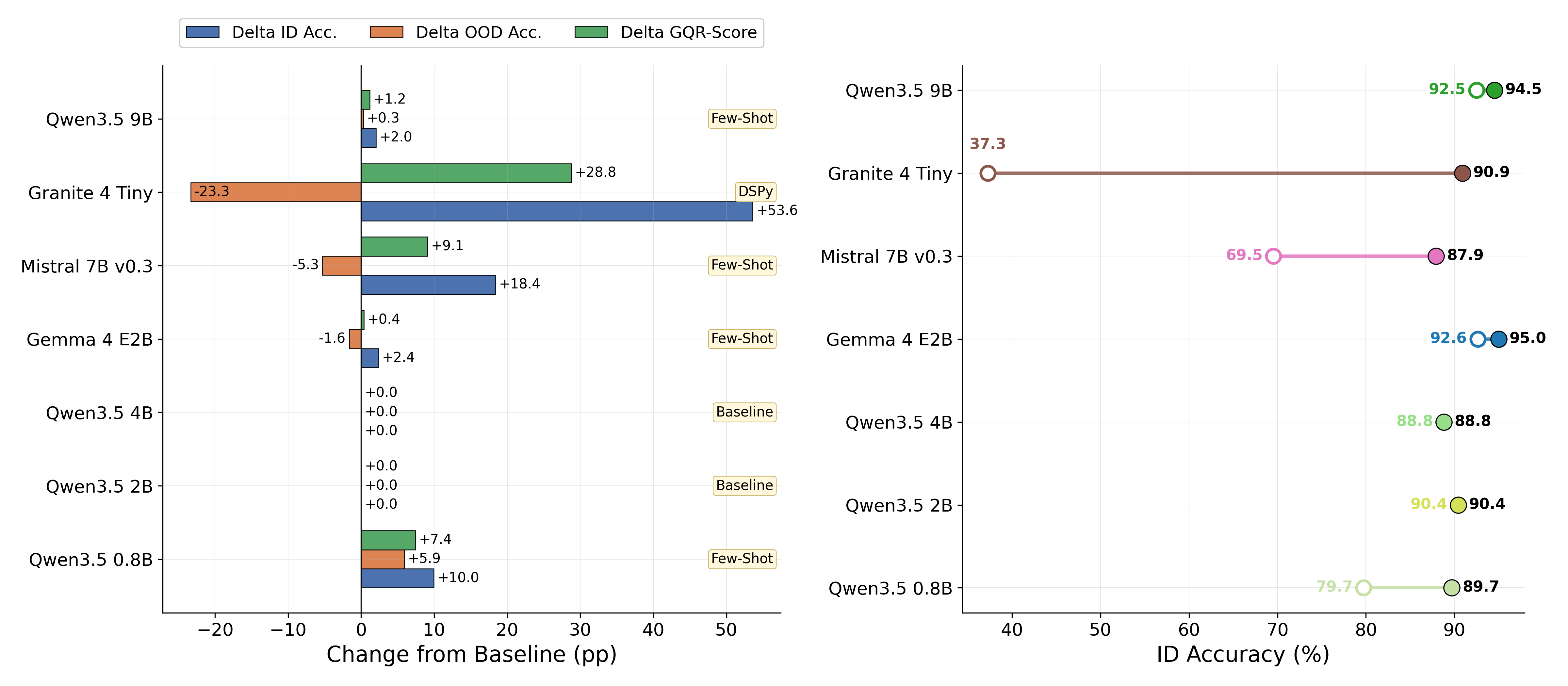}
\caption{Decomposition of GQR-Score gains from prompt optimization. The left panel shows the change in ID accuracy (blue), OOD accuracy (orange), and GQR-Score (green) relative to standard prompting for each model's best strategy (annotated). The right panel shows absolute ID accuracy before (hollow) and after (filled) optimization, with connecting lines indicating the magnitude of ID recovery. The figure shows that prompt optimization mainly helps by repairing ID routing, not by further improving already high OOD rejection.}
\label{fig:optimization-gains}
\end{figure}

Figure~\ref{fig:optimization-gains} shows where the gains come from. Prompt optimization does not mainly improve OOD rejection, which was already high for most models. Instead, it recovers ID routing. Granite~4~Tiny rises from 37.29\% to 94.88\% ID accuracy under Few-Shot, and Mistral~7B rises from 69.53\% to 87.93\%. The exception is Qwen3.5~0.8B, whose standard-prompting OOD accuracy was unusually low (48.55\%) and therefore also improves on OOD.

\paragraph{Effect of Few-shot prompting} Few-shot constitutes the single most effective prompt-only strategy~\cite{b2}. Under Few-Shot optimization, Qwen3.5~9B reaches 95.74, within 0.3 points of the best unoptimized model (Gemma~3~27B at 96.01); Gemma~4~E2B follows at 94.19, Qwen3.5~4B at 92.51, and Mistral~7B at 90.87.

The underlying mechanism is ID recovery through exemplar-based format demonstration. Granite~4~Tiny improves from 37.29\% to 94.88\% ID accuracy (+57.6\,pp), and Mistral~7B from 69.53\% to 87.93\% (+18.4\,pp). These models did not lack domain knowledge. Their baseline OOD accuracy already exceeded 99\%, but they failed to commit to an ID label under the strict output schema. In-context exemplars resolve this by demonstrating the expected format of a correct routing decision~\cite{b30,b31,b36,b37}. Both models pay an OOD cost. Granite~4~Tiny incurs a $-$32.6\,pp OOD penalty (from 99.75\% to 67.18\%), and Mistral~7B incurs a smaller $-$5.3\,pp OOD penalty (from 99.29\% to 94.02\%). The net Few-Shot GQR gains of +24.4 (Granite) and +9.1 (Mistral) still justify the exchange, as the standard-prompting routers were effectively non-functional on the ID side; for Granite~4~Tiny, the DSPy baseline further improves the trade-off to 83.05 GQR.

Two models illustrate where few-shot optimization breaks down. Qwen3.5~2B drops from a baseline GQR of 90.58 to 70.43 under Few-Shot. ID accuracy rises to 93.78\%, but OOD accuracy falls to 56.40\%, indicating that the exemplars overcorrect and the model ceases to reject out-of-distribution inputs. Qwen3.5~0.8B attains a Few-Shot GQR of only 67.79, suggesting that a 0.8B parameter budget is fragile for reliable four-way classification under exemplar-based prompts.

\paragraph{Effect of GEPA Refinement} GEPA refinement~\cite{b3} does not yield the best GQR-Score for any of the seven models evaluated. For Qwen3.5~9B, GEPA reaches 95.58, above the standard-prompting baseline (94.55) but below Few-Shot (95.74). For Qwen3.5~0.8B, GEPA actually underperforms both Few-Shot (67.79) and the standard-prompting baseline (60.35), reaching only 48.92. For the remaining models, GEPA generally lands between the standard baseline and Few-Shot: Gemma~4~E2B drops from 94.19 (Few-Shot) to 89.88, and Mistral~7B drops from 90.87 to 82.94. Granite~4~Tiny is the one model where GEPA (81.58) exceeds Few-Shot (78.66), though the DSPy baseline (83.05) remains the best strategy for that model.

This pattern is interpretable. For mid-strength models that already achieved a strong Few-Shot prompt, the instruction-optimization landscape is brittle. Minor perturbations either preserve routing semantics (no effect) or shift the OOD/ID balance enough to harm GQR (degradation), leaving little room for improvement through instruction editing alone. The bootstrapped few-shot prompt already sits at a robust local optimum that evolutionary search cannot easily escape within this budget, which separates constrained classification from the open-ended generation tasks where GEPA has shown larger gains~\cite{b3,b29}.

\subsection{Latency}

\begin{figure}[t]
\centering
\includegraphics[width=\linewidth]{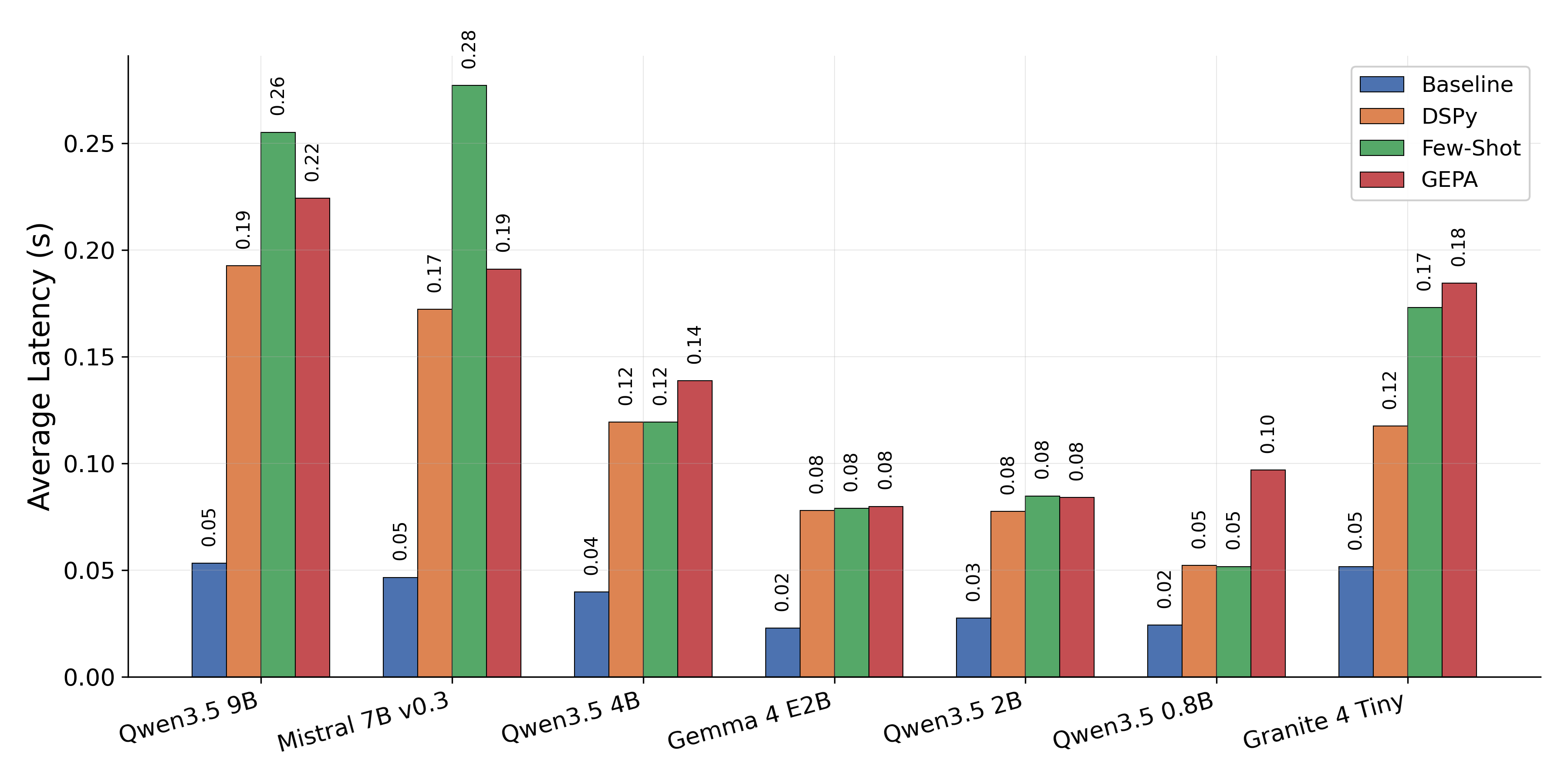}
\caption{Average router inference latency by prompting strategy. Few-Shot increases latency primarily because in-context exemplars lengthen the prompt. The figure clarifies the central deployment trade-off, showing that prompt optimization can recover routing accuracy, but the longer prompt must still fit the latency budget of a router that runs on every request.}
\label{fig:latency-by-approach}
\end{figure}

Prompt optimization increases inference latency because optimized prompts are longer (Figure~\ref{fig:latency-by-approach}). Under standard prompting, all seven optimized models run between 0.02\,s (Gemma~4~E2B) and 0.05\,s (Qwen3.5~9B) per query, well within interactive routing budgets. After optimization, latencies cluster between 0.05\,s and 0.28\,s across DSPy, Few-Shot, and GEPA strategies, roughly a 2$\times$ to 6$\times$ increase. All three strategies produce prompts of similar length and therefore yield comparable latencies for most models; differences across strategies for a given model are small relative to differences across models. The non-LLM baselines (fastText, WideMLP) remain two to three orders of magnitude faster~\cite{b1}, so whether the GQR gain justifies the added latency depends on the error-cost asymmetry of the deployment.

\section{Discussion}

Taken together, the results support a number of recommendations for practice: Compact models do not automatically serve as effective guarded routers, but mid-scale SLMs become strong once the prompt teaches them the required format. Under Few-Shot, Qwen3.5~9B (95.74) comes within 0.3 GQR points of the best unoptimized model, Gemma~3~27B (96.01), while running at lower latency (Figure~\ref{fig:tradeoff}), with Gemma~4~E2B (94.19) and Qwen3.5~4B (92.51) following. Nevertheless, this does not mean that every SLM should replace a classical classifier. WideMLP with confidence thresholding reaches 87.74~\cite{b1} and runs two to three orders of magnitude faster. The useful SLM regime is therefore the middle range, where models are large enough to follow the format but small enough to remain practical routers~\cite{b19,b20}.

The bottleneck for most of these models is not OOD detection but ID commitment, and within ID commitment, two qualitatively different failure modes appear: Mistral~7B, Granite~4~Tiny, and Llama~3.2~3B reject OOD queries reliably (over 99\% OOD accuracy) but under-commit on ID, defaulting to the rejection label even for valid domain queries. This is a partial-commit failure, and it is largely recoverable through prompting: optimization lifts the ID accuracy of Granite~4~Tiny from 37.29\% to 94.88\% and that of Mistral~7B from 69.53\% to 87.93\% without weight updates. Larger models such as Phi-4~14B, Gemma~3~12B, Qwen3.5~9B, and GPT-OSS~20B exhibit neither pattern, which suggests that GQR is not inherently a trade-off between ID and OOD accuracy. The trade-off appears specifically when a model is capable enough to sense the domain but does not commit to one of the four allowed labels~\cite{b36,b37}.

No single optimization strategy dominates. Few-Shot yields the best GQR for four of the seven optimized models, the DSPy baseline is the best strategy for Granite~4~Tiny (54.29~$\rightarrow$~83.05), and standard prompting already beats both optimized variants for Qwen3.5~4B and Qwen3.5~2B. GEPA never produces the best score for any model in this set, although it does improve over standard prompting for Qwen3.5~9B (94.55~$\rightarrow$~95.58) and Granite~4~Tiny (54.29~$\rightarrow$~81.58). Once a few-shot prompt has taught the four-label format, the instruction-optimization landscape is brittle: small edits either preserve the same behavior or break the format. This differs from open-ended generation, in which reflective refinement has more freedom to improve the answer~\cite{b3,b29}.

Prompt-only optimization is also limited at the smaller end of the scale. Qwen3.5~0.8B reaches only 67.79 even under Few-Shot prompting, and the still smaller Gemma~3~1B (42.22) and Gemma~3~270M (8.30), which sit below the 0.8B threshold and were not part of the seven-model optimization subset (Section~\ref{subsec:models}), already collapse under standard prompting (Listing~\ref{lst:baseline-router-prompt}), so adding exemplars is unlikely to rescue them. Conversely, a well-calibrated standard-prompting score can be damaged by exemplars when there is little to fix, as Qwen3.5~2B illustrates.

The four-label schema is easy to deploy, but it conflates at least three failure modes that Table~\ref{tab:model-ood-results-full-big} alone cannot separate: a model may lack the domain knowledge, may know the domain but fail to follow the format, or may refuse on principle. The single GQR-Score collapses these into one number, which makes it difficult to choose between weight-level adaptation, prompt repair, and schema redesign for any given model. The same observation applies at the column level: Web~Q and OLID are the hardest rejection splits under standard prompting (Table~\ref{tab:model-ood-results-full-big}) and remain the hardest after optimization, but the schema does not indicate whether this reflects a domain-boundary problem or a residual format-following problem.

\paragraph{Future work} should investigate weight-level adaptation, in particular, Low-Rank Adaptation (LoRA) or Quantized LoRA (QLoRA) fine-tuning combined with DSPy optimized signatures, to determine whether the format-following bottleneck identified here can be resolved below the prompt layer. Compared with prompt-only methods, lightweight adaptation would require a training and deployment step, but it may provide a more stable way to bind compact models to the allowed label set and reduce the brittle overcorrection observed for Qwen3.5~2B and Qwen3.5~0.8B.

\section{Conclusion}

We benchmarked 22 open-weight models on GQR-Bench and found a clear pattern. The best mid-scale SLMs can approach much larger models, but weaker models often fail in a specific way. They reject OOD inputs well, yet do not reliably choose the right ID label. Few-shot prompting repairs much of this format-following problem. Mistral~7B rises from 81.79 to 90.87 GQR-Score, Granite~4~Tiny rises from 54.29 to 83.05 under the DSPy baseline, and Few-Shot brings Qwen3.5~9B to 95.74, just below the strongest unoptimized model, Gemma~3~27B at 96.01. Compact open-weight SLMs can therefore be practical guarded routers in the tested GQR-Bench setting, but broader deployment still requires testing other schemas, label granularities, and lightweight adaptation methods.

\section{Limitations}
\label{sec:limitations}

All results are single-run point estimates on a fixed two-GPU configuration (2$\times$ RTX~4090); no run-to-run variance is characterized, and the shape of the Pareto frontier is hardware-dependent~\cite{b18}. GEPA uses gpt-5.4 as the teacher model, and the effectiveness of evolutionary refinement with weaker or open-weight teachers remains untested. The analysis is also tied to the GQR-Bench schema, with three ID domains and one catch-all OOD label. Other routing deployments may require finer label granularity, hierarchical abstention labels, or different output formats, which could change the balance between prompt-following failures and true domain errors. Finally, we do not evaluate LoRA, QLoRA, or other lightweight weight-adaptation methods; these methods may mitigate the format-following bottleneck more directly than prompt edits, but they introduce additional training and deployment costs that fall outside this study.

\section{Ethical Considerations}

Over-rejection by a guarded router denies legitimate users access to legal, financial, or medical assistance, a failure mode that disproportionately affects queries sharing surface vocabulary with sensitive content. The high-OOD, low-ID profiles of Mistral~7B and Granite~4~Tiny illustrate this risk in practice. A router deployed with either model under standard prompting would block the majority of valid requests. We also note a dual-use concern, since a well-calibrated guarded router can function as a content gate, and narrowing the ID taxonomy without disclosure could enable selective censorship.

\begin{credits}
\subsubsection{\ackname} This work was supported by the Science Grant Agency -- project VEGA 1/0300/25.
This research was further supported in part by the MIST project, funded by the Novo Nordisk Foundation under grant reference number NNF25OC0103204.

\subsubsection{\discintname} The authors have no competing interests to declare that are relevant to the content of this article.
\end{credits}

\bibliographystyle{splncs04}
\bibliography{mybibliography}

@inproceedings{b1,
  author    = {Richard {\v{S}}l{\'e}her and William Brach and Tom{\'a}{\v{s}} Sloboda and Karel Ko{\v{s}}{\v{t}}{\'a}{\v{l}} and Luk{\'a}{\v{s}} Galke},
  title     = {Guarded Query Routing for Large Language Models},
  booktitle = {ECAI},
  year      = {2025}
}

@inproceedings{b2,
  author    = {Omar Khattab and others},
  title     = {{DSPy}: Compiling Declarative Language Model Calls into State-of-the-Art Pipelines},
  booktitle = {ICLR},
  year      = {2024}
}

@article{dekoninck2024unified,
  title={A unified approach to routing and cascading for llms},
  author={Dekoninck, Jasper and Baader, Maximilian and Vechev, Martin},
  journal={arXiv preprint arXiv:2410.10347},
  year={2024}
}

@misc{b3,
  author       = {Amit Agrawal and others},
  title        = {{GEPA}: Reflective Prompt Evolution Can Outperform Reinforcement Learning},
  year         = {2025},
  eprint       = {2507.19457},
  archivePrefix= {arXiv},
  url          = {https://arxiv.org/pdf/2507.19457.pdf}
}

@misc{b4,
  author       = {An Yang and others},
  title        = {Qwen3 Technical Report},
  year         = {2025},
  eprint       = {2505.09388},
  archivePrefix= {arXiv},
  url          = {https://arxiv.org/pdf/2505.09388.pdf}
}

@misc{b5,
  author       = {Angela Grattafiori and others},
  title        = {The Llama 3 Herd of Models},
  year         = {2024},
  eprint       = {2407.21783},
  archivePrefix= {arXiv},
  url          = {https://arxiv.org/pdf/2407.21783.pdf}
}

@misc{b6,
  author       = {Mohamed Abdin and others},
  title        = {Phi-4 Technical Report},
  year         = {2024},
  eprint       = {2412.08905},
  archivePrefix= {arXiv},
  url          = {https://arxiv.org/pdf/2412.08905.pdf}
}

@misc{b7,
  author       = {{Gemma Team}},
  title        = {Gemma 3 Technical Report},
  year         = {2025},
  eprint       = {2503.19786},
  archivePrefix= {arXiv},
  url          = {https://arxiv.org/pdf/2503.19786.pdf}
}

@inproceedings{b9,
  author    = {Myong Chol Jung and He Zhao and Joanna Dipnall and Belinda Gabbe and Lan Du},
  title     = {Enhancing Near Out-of-Distribution Detection in Prompt Learning: Maximum Gains, Minimal Costs},
  booktitle = {ICLR},
  year      = {2025}
}

@inproceedings{b10,
  author    = {Ruiyao Xu and Kaize Ding},
  title     = {Large Language Models for Anomaly and Out-of-Distribution Detection: A Survey},
  booktitle = {Findings of the Association for Computational Linguistics: NAACL},
  year      = {2025}
}

@inproceedings{b11,
  author    = {Javid Lakha and Minlan Yu and Rana Shahout},
  title     = {Faster, Cheaper, Just as Good: Cost- and Latency-Constrained Routing for LLMs},
  booktitle = {ICLR},
  year      = {2025}
}

@article{b12,
  author  = {F. Wang and others},
  title   = {A Comprehensive Survey of Small Language Models in the Era of Large Language Models: Techniques, Enhancements, Applications, Collaboration with LLMs, and Trustworthiness},
  journal = {ACM Transactions on Intelligent Systems and Technology},
  year    = {2025}
}

@misc{b13,
  author       = {Albert Q. Jiang and others},
  title        = {Mistral 7B},
  year         = {2023},
  eprint       = {2310.06825},
  archivePrefix= {arXiv},
  url          = {https://arxiv.org/abs/2310.06825}
}

@misc{b14,
  author       = {{IBM Granite Team}},
  title        = {Granite 3.3 Language Models},
  year         = {2025},
  url          = {https://huggingface.co/ibm-granite/granite-3.3-8b-instruct}
}

@misc{b18,
  author = {{Ollama}},
  title  = {Ollama},
  year   = {2025},
  url    = {https://github.com/ollama/ollama}
}

@article{b19,
  author  = {Lingjiao Chen and Matei Zaharia and James Zou},
  title   = {FrugalGPT: How to Use Large Language Models While Reducing Cost and Improving Performance},
  journal = {Transactions on Machine Learning Research},
  year    = {2023}
}

@inproceedings{b20,
  author    = {Jonas Dekoninck and Michael Baader and Martin Vechev},
  title     = {A Unified Approach to Routing and Cascading for {LLMs}},
  booktitle = {ICML},
  year      = {2025},
  pages     = {12987--13010}
}

@inproceedings{b21,
  author    = {Tal Shnitzer and others},
  title     = {Large Language Model Routing with Benchmark Datasets},
  booktitle = {CoLM},
  year      = {2024}
}

@inproceedings{b22,
  author    = {Dan Hendrycks and Kevin Gimpel},
  title     = {A Baseline for Detecting Misclassified and Out-of-Distribution Examples in Neural Networks},
  booktitle = {ICLR},
  year      = {2017}
}

@inproceedings{b23,
  author    = {Shiyu Liang and Yixuan Li and R. Srikant},
  title     = {Enhancing the Reliability of Out-of-Distribution Image Detection in Neural Networks},
  booktitle = {ICLR},
  year      = {2018}
}

@inproceedings{b24,
  author    = {Caterina Tomani and Kamalika Chaudhuri and Ivan Evtimov and Daniel Cremers and Mohamed Ibrahim},
  title     = {Uncertainty-Based Abstention in Large Language Models Improves Safety and Reduces Hallucinations},
  booktitle = {ICLR},
  year      = {2025}
}

@inproceedings{b25,
  author    = {S. Tayebati and D. Kumar and N. Darabi and D. Jayasuriya and R. Krishnan and A. R. Trivedi},
  title     = {{CAP}: Conformalized Abstention Policies for Context-Adaptive Risk Management for {LLMs} and {VLMs}},
  booktitle = {Asian Conference on Machine Learning (ACML)},
  year      = {2025}
}

@misc{b26,
  author       = {S. Schulhoff and others},
  title        = {The Prompt Report: A Systematic Survey of Prompting Techniques},
  year         = {2024},
  eprint       = {2406.06608},
  archivePrefix= {arXiv},
  url          = {https://arxiv.org/pdf/2406.06608.pdf}
}

@inproceedings{b27,
  author    = {Taylor Shin and Yashar Razeghi and Robert L. Logan IV and Eric Wallace and Sameer Singh},
  title     = {AutoPrompt: Eliciting Knowledge from Language Models with Automatically Generated Prompts},
  booktitle = {EMNLP},
  year      = {2020}
}

@inproceedings{b28,
  author    = {Brian Lester and Rami Al-Rfou and Noah Constant},
  title     = {The Power of Scale for Parameter-Efficient Prompt Tuning},
  booktitle = {EMNLP},
  year      = {2021},
  pages     = {3045--3059}
}

@inproceedings{b29,
  author    = {Aman Madaan and others},
  title     = {Self-Refine: Iterative Refinement with Self-Feedback},
  booktitle = {Advances in Neural Information Processing Systems (NeurIPS)},
  year      = {2023}
}

@inproceedings{b30,
  author    = {Tom B. Brown and others},
  title     = {Language Models are Few-Shot Learners},
  booktitle = {Advances in Neural Information Processing Systems (NeurIPS)},
  year      = {2020}
}

@inproceedings{b31,
  author    = {Timo Schick and Hinrich Sch{\"u}tze},
  title     = {Exploiting Cloze Questions for Few-Shot Text Classification and Natural Language Inference},
  booktitle = {EACL},
  year      = {2021}
}

@inproceedings{b32,
 author = {Zhao, Eric and Awasthi, Pranjal and Chen, Zhengdao and Gollapudi, Sreenivas and Delling, Daniel},
 booktitle = {Advances in Neural Information Processing Systems},
 doi = {10.52202/079017-0323},
 editor = {A. Globerson and others},
 pages = {10060--10087},
 publisher = {Curran Associates, Inc.},
 title = {Semantic Routing via Autoregressive Modeling},
 volume = {37},
 year = {2024}
}

@misc{b33,
      title={gpt-oss-120b \& gpt-oss-20b Model Card}, 
      author={{{OpenAI} and S. Agarwal and others}},
      year={2025},
      eprint={2508.10925},
      archivePrefix={arXiv},
      primaryClass={cs.CL},
      url={https://arxiv.org/abs/2508.10925}, 
}

@article{b34,
  title={xRouter: Training Cost-Aware LLMs Orchestration System via Reinforcement Learning},
  author={Qian and others},
  journal={arXiv preprint arXiv:2510.08439},
  year={2025}
}

@misc{b35,
  author       = {Hakan Inan and others},
  title        = {Llama Guard: LLM-based Input-Output Safeguard for Human-AI Conversations},
  year         = {2023},
  eprint       = {2312.06674},
  archivePrefix= {arXiv},
  url          = {https://arxiv.org/abs/2312.06674}
}

@misc{b36,
  author       = {Jeffrey Zhou and Tianjian Lu and Swaroop Mishra and Siddhartha Brahma and Sujoy Basu and Yi Luan and Denny Zhou and Le Hou},
  title        = {Instruction-Following Evaluation for Large Language Models},
  year         = {2023},
  eprint       = {2311.07911},
  archivePrefix= {arXiv},
  url          = {https://arxiv.org/abs/2311.07911}
}

@misc{b37,
  author       = {Brandon T. Willard and R{\'e}mi Louf},
  title        = {Efficient Guided Generation for Large Language Models},
  year         = {2023},
  eprint       = {2307.09702},
  archivePrefix= {arXiv},
  url          = {https://arxiv.org/abs/2307.09702}
}

@inproceedings{b38,
  author    = {Weitang Liu and Xiaoyun Wang and John D. Owens and Yixuan Li},
  title     = {Energy-based Out-of-distribution Detection},
  booktitle = {Advances in Neural Information Processing Systems (NeurIPS)},
  year      = {2020}
}

@misc{b39,
  author       = {Isaac Ong and Amjad Almahairi and Vincent Wu and Wei-Lin Chiang and Tianhao Wu and Joseph E. Gonzalez and M. Waleed Kadous and Ion Stoica},
  title        = {{RouteLLM}: Learning to Route {LLMs} with Preference Data},
  year         = {2024},
  eprint       = {2406.18665},
  archivePrefix= {arXiv},
  url          = {https://arxiv.org/abs/2406.18665}
}

@inproceedings{b40,
  author    = {Traian Rebedea and Razvan Dinu and Makesh Sreedhar and Christopher Parisien and Jonathan Cohen},
  title     = {{NeMo Guardrails}: A Toolkit for Controllable and Safe {LLM} Applications with Programmable Rails},
  booktitle = {EMNLP: System Demonstrations},
  year      = {2023}
}

@inproceedings{b41,
  author    = {Krista Opsahl-Ong and Michael J. Ryan and Josh Purtell and David Broman and Christopher Potts and Matei Zaharia and Omar Khattab},
  title     = {Optimizing Instructions and Demonstrations for Multi-Stage Language Model Programs},
  booktitle = {EMNLP},
  year      = {2024}
}

@inproceedings{galke-scherp-2022-bag,
    title = "Bag-of-Words vs. Graph vs. Sequence in Text Classification: Questioning the Necessity of Text-Graphs and the Surprising Strength of a Wide {MLP}",
    author = "Galke, Lukas  and
      Scherp, Ansgar",
    no_editor = "Muresan, Smaranda  and
      Nakov, Preslav  and
      Villavicencio, Aline",
    booktitle = "ACL",
    no_month = may,
    year = "2022",
    no_address = "Dublin, Ireland",
    no_publisher = "Association for Computational Linguistics",
    no_url = "https://aclanthology.org/2022.acl-long.279/",
    doi = "10.18653/v1/2022.acl-long.279",
    pages = "4038--4051"
}

@article{bojanowski2017enriching,
  title={Enriching word vectors with subword information},
  author={Bojanowski, Piotr and Grave, Edouard and Joulin, Armand and Mikolov, Tomas},
  journal={Transactions of the association for computational linguistics},
  volume={5},
  pages={135--146},
  year={2017},
  publisher={MIT Press One Rogers Street, Cambridge, MA 02142-1209, USA journals-info~…}
}

@inproceedings{joulin2017bag,
  title={Bag of tricks for efficient text classification},
  author={Joulin, Armand and Grave, Edouard and Bojanowski, Piotr and Mikolov, Tom{\'a}{\v{s}}},
  booktitle={EACL: Short papers},
  pages={427--431},
  year={2017}
}

@article{galke2022we,
  title={Are we really making much progress in text classification? a comparative review},
  author={Galke, Lukas and Diera, Andor and Lin, Bao Xin and Khera, Bhakti and Meuser, Tim and Singhal, Tushar and Karl, Fabian and Scherp, Ansgar},
  journal={arXiv preprint arXiv:2204.03954},
  year={2022}
}

@inproceedings{borkan2019jigsaw,
  author    = {Daniel Borkan and Lucas Dixon and Jeffrey Sorensen and Nithum Thain and Lucy Vasserman},
  title     = {Nuanced Metrics for Measuring Unintended Bias with Real Data for Text Classification},
  booktitle = {Companion Proceedings of The 2019 World Wide Web Conference (WWW)},
  year      = {2019},
  pages     = {491--500},
  doi       = {10.1145/3308560.3317593}
}

@inproceedings{zampieri2019olid,
  author    = {Marcos Zampieri and Shervin Malmasi and Preslav Nakov and Sara Rosenthal and Noura Farra and Ritesh Kumar},
  title     = {Predicting the Type and Target of Offensive Posts in Social Media},
  booktitle = {Proceedings of the Conference of the North American Chapter of the Association for Computational Linguistics: Human Language Technologies (NAACL-HLT)},
  year      = {2019},
  pages     = {1415--1420}
}

@inproceedings{mathew2021hatexplain,
  author    = {Binny Mathew and Punyajoy Saha and Seid Muhie Yimam and Chris Biemann and Pawan Goyal and Animesh Mukherjee},
  title     = {{HateXplain}: A Benchmark Dataset for Explainable Hate Speech Detection},
  booktitle = {Proceedings of the AAAI Conference on Artificial Intelligence},
  volume    = {35},
  number    = {17},
  pages     = {14867--14875},
  year      = {2021}
}

@inproceedings{sigurbergsson2020dkhate,
  author    = {Gudbjartur Ingi Sigurbergsson and Leon Derczynski},
  title     = {Offensive Language and Hate Speech Detection for {Danish}},
  booktitle = {Proceedings of the 12th Language Resources and Evaluation Conference (LREC)},
  year      = {2020},
  pages     = {3498--3508}
}

@inproceedings{hladek2023tukesk,
  author    = {Daniel Hl{\'a}dek and J{\'a}n Sta{\v{s}} and Mat{\'u}{\v{s}} Pleva and Yuriy Bobr{\'y}ek},
  title     = {Slovak Dataset for Hate Speech Detection},
  booktitle = {Proceedings of the International Conference on Emerging eLearning Technologies and Applications (ICETA)},
  year      = {2023},
  organization = {IEEE}
}

\end{document}